\documentstyle[12pt,epsfig]{article}
\newlength{\dinwidth}
\newlength{\dinmargin}
\newcommand{\ba}{\begin{array}}
\newcommand{\ea}{\end{array}}
\newcommand{\beq}{\begin{equation}}
\newcommand{\eeq}{\end{equation}}
\newcommand{\bea}{\begin{eqnarray}}
\newcommand{\eea}{\end{eqnarray}}

\def\S{{\bf S}}

\def\bce{\begin{center}}
\def\ece{\end{center}}

\def\nonu{\nonumber}

\def\pa{\partial}
\def\al{\alpha}
\def\be{\beta}

\def\Ga{\Gamma}

\def\De{\Delta}

\def\R{{\bf R}}

\def\S{{\bf S}}

\begin{document}
\thispagestyle{empty}
\addtocounter{page}{-1}
\begin{flushright}
{\tt hep-th/0206029}\\
\end{flushright}
\vspace*{1.3cm}
\centerline{\Large \bf Penrose Limit of $AdS_4 \times V_{5,2}$
and \Large \bf  Operators with Large $R$ Charge}
\vspace*{1.5cm} 
\centerline{ \bf Changhyun Ahn
}
\vspace*{1.0cm}
\centerline{\it 
Department of Physics,}
\vskip0.3cm 
\centerline{  \it Kyungpook National University,}
\vskip0.3cm
\centerline{  \it  Taegu 702-701, Korea }
\vspace*{0.3cm}
\centerline{\tt ahn@knu.ac.kr 
}
\vskip2cm
\centerline{\bf  abstract}
\vspace*{0.5cm}

We consider M-theory on $AdS_4 \times V_{5,2}$ where
$V_{5,2}= SO(5)/SO(3)$ is a Stiefel manifold. 
We construct a Penrose limit of $AdS_4 \times V_{5,2}$ that 
provides the pp-wave geometry. 
There exists a subsector of 
three dimensional ${\cal N}=2$ dual gauge theory, by taking
both the conformal dimension and $R$ charge large with the finiteness 
of their difference, which 
has enhanced ${\cal N}=8$ maximal supersymmetry.
We identify operators in the ${\cal N}=2$ gauge theory with
supergravity KK excitations in the pp-wave geometry and 
describe how the gauge theory operators made out of
chiral field of conformal dimension 1/3 fall into 
${\cal N}=8$ supermultiplets.

\vspace*{4.0cm}

%\begin{flushleft}
%{Aug., 1999}\\
%\end{flushleft}
%\centerline{\bf DRAFT VERSION}
%\centerline{Submitted to Nuclear Physics B}
\baselineskip=18pt
\newpage

%%%%%%%%%%%%%%%%%%%%%%%%%%%%%%%%%%%%%%%%%%%%%%%%%%%%%%%%%%%%%%%%%%%%%%%%%%%%%
\section{Introduction}
%%%%%%%%%%%%%%%%%%%%%%%%%%%%%%%%%%%%%%%%%%%%%%%%%%%%%%%%%%%%%%%%%%%%%%%%%%%%%

The large $N$ limit of a subsector of $d=4, {\cal N}=4$ $SU(N)$
supersymmetric gauge theory is dual \cite{bmn} to type IIB string theory in the
pp-wave background \cite{bfhpetal,bfhpetal1}. 
This subspace of the gauge theory
is described by string theory in the pp-wave background. By taking
a scale limit of the geometry near a null geodesic in $AdS_5 \times \S^5$,
it gives rise to the appropriate subspace of the gauge theory. The operators
with large $R$-charge in the subsector of ${\cal N}=4$ $SU(N)$ gauge theory
were identified with the stringy states in the pp-wave background.
There exists a particularly interesting model by replacing $\S^5$ with
a five-dimensional manifold, $T^{1,1}$ with lower supersymmetry. 
It was found that the Penrose limit of $AdS_5 \times T^{1,1}$ provides
pp-wave geometry of $AdS_5 \times \S^5$ \cite{ikm,go,zs}. 
Using AdS/CFT correspondence, one
can identify gauge theory operators with large $R$-charge with the stringy
excitations in the pp-wave geometry. Moreover, the maximal ${\cal N}=4$
multiplet structure hidden in the ${\cal N}=1$ gauge theory can be 
predicted from both a chiral operator and semi-conserved operator with large 
$R$-charge.  There are many papers \cite{etal} on the work of \cite{bmn}.  
   
It is natural to think about the subsector of ${\cal N}=2$
gauge theory in $d=3$ in the context of $AdS_4 \times {\bf X}^7$ where
${\bf X}^7$ is an Einstein seven manifold. 
Recently the operators with large $R$-charge in the boundary field 
theory were obtained from the complete spectrums of 11-dimensional KK 
compactifications on $AdS_4 \times Q^{1,1,1}$ \cite{ahn02} 
and $AdS_4 \times M^{1,1,1}$ \cite{ahn02-1} in pp-wave limit. In old days,
all the supersymmetric ${\cal N}=2$ 
homogeneous manifolds were classified in \cite{crw}.
There exist only three ${\cal N}=2$ theories and they are $Q^{1,1,1}, 
M^{1,1,1}$ and $V_{5,2}$. The isometry of these manifolds corresponds to
the global symmetry of the dual SCFT including $U(1)_R$ symmetry of
${\cal N}=2$ supersymmetry.      

In this paper, we consider a similar duality
that is present between a certain three dimensional ${\cal N}=2$
gauge theory and 11-dimensional supergravity theory
in a pp-wave background with the same spirit as in 
\cite{ahn02,ahn02-1,go,ikm,zs}.
This is a continuation of previous considerations \cite{ahn02,ahn02-1}. 
We describe this duality by taking a scaling limit of the duality
between 11-dimensional supergravity on $AdS_4 \times V_{5,2}$
where $V_{5,2}$ was found in \cite{crw} and three dimensional
superconformal field theory. The boundary theory is a gauge theory
with gauge group $USp(2N) \times O(2N-1)$ with chiral fields
$S^i$ transforming in the $(2N,2N-1)$ color representation and
transforming in the spinor representation of the flavor group
$SO(5)$. 
The complete analysis on the spectrum of
$AdS_4 \times V_{5,2}$ was found in \cite{cddf}. 
This gives the theory that lives on $N$ M2-branes 
at the conical singularity of a Calabi-Yau four-fold. 
The scaling limit is obtained by considering the geometry near a null 
geodesic carrying large angular momentum in the $U(1)_R$
isometry of the $V_{5,2}$ space which is dual to the $U(1)_R$ R-symmetry
in the ${\cal N}=2$ superconformal field theory.

In section 2, we consider the scaling limit around
a null geodesic in $AdS_4 \times V_{5,2}$ from the explicit
metric of $V_{5,2}$ given in terms of angular variables \cite{bh}
and obtain a pp-wave 
background.
In section 3, we identify supergravity excitations in the Penrose limit
with gauge theory operators. What we observed is the presence of
a semi-conserved field in the long vector multiplet II propagating
in the $AdS_4$ bulk. 
In section 4, we summarize our results.

%%%%%%%%%%%%%%%%%%%%%%%%%%%%%%%%%%%%%%%%%%%%%%%%%%%%%%%%%%%%%%%%%%%%%%%%%%%%%
\section{Penrose Limit of $AdS_4 \times V_{5,2}$}
%%%%%%%%%%%%%%%%%%%%%%%%%%%%%%%%%%%%%%%%%%%%%%%%%%%%%%%%%%%%%%%%%%%%%%%%%%%%%

Let us start with the supergravity solution dual to
the ${\cal N}=2$ superconformal field theory \cite{cddf}.
By putting a large number of $N$ coincident M2-branes at the
conifold singularity and taking the near horizon limit,
the metric becomes  that \cite{bh} of 
$AdS_4 \times V_{5,2}$(See also \cite{cveticetal})
\bea
ds_{11}^2 =ds_{AdS_4}^2 +ds_{V_{5,2}}^2,
\label{metric}
\eea
where
\bea
ds_{AdS_4}^2 & = & L^2 \left( -\cosh^2 \rho \; d t^2 +d \rho^2 + 
\sinh^2 \rho \; d \Omega_2^2 \right),
\nonu \\
ds_{V_{5,2}}^2 & = & 
\frac{9L^2}{16}  \left[ d \psi + \frac{1}{2} \cos \al 
\left(  d \be - \sum_{i=1}^2 \cos \theta_i d \phi_i \right) \right]^2 +
\frac{3L^2}{8} d \al^2 \nonu \\
&& +\frac{3L^2}{32}  \sin^2 \al
\left( d \be - \sum_{i=1}^2 \cos \theta_i d \phi_i \right)^2
+ \frac{3L^2}{32} \left( 1 + \cos^2 \al \right)  
\sum_{i=1}^2 \left( \sin^2 \theta_i d \phi_i^2 + d \theta_i^2 \right)
 \nonu \\
& & + \frac{3L^2}{16} \sin^2 \al \cos \be \sin \theta_1 \sin \theta_2 d \phi_1
d \phi_2 - \frac{3L^2}{16} \sin^2 \al \cos \be d \theta_1 d \theta_2 \nonu \\
&& + \frac{3L^2}{16} \sin^2 \al \sin \be \left( \sin \theta_1 d \phi_1 
d \theta_2 + \sin \theta_2 d \phi_2 d \theta_1 \right),
\label{metric1}
\eea
where $ d \Omega_2$ is the volume form of a unit $\S^2$
and the curvature radius 
$L$ of $AdS_4$ is given by
$(2L)^6= 32 \pi^2 \ell_p^6 N$. 
%Topologically $M^{1,1,1}$ 
%s a nontrivial $U(1)$ bundle
%over ${\bf CP}^2 \times \S^2$. 
The spherical coordinates $(\theta_i, \phi_i)$ parametrize
two sphere $\S^2_i$, as usual, and
the angles vary over the ranges, $0 \leq \theta_i \leq  \pi,
0 \leq \phi_i, \psi \leq 2 \pi, 0 \leq \be \leq 4 \pi$ and $ 0 \leq \al 
\leq \pi/2$.
The $SO(5) \times U(1) $ isometry
 group of $V_{5,2}$ consists of $SO(5)
 $ global symmetry and $U(1)_R$ symmetry of the dual 
superconformal field theory of \cite{cddf}.

Let us make a scaling limit around a null geodesic in 
 $AdS_4 \times V_{5,2}$ that rotates along the $\psi$ coordinate
of $V_{5,2}$ whose shift symmetry corresponds to
the $U(1)_R$ symmetry in the dual superconformal field theory.  
Let us introduce
coordinates which label the geodesic   
\bea
x^{+}  =  \frac{1}{2} \left[ t +\frac{3}{4} \left( \psi + 
\frac{1}{2} \be -\frac{1}{2} \phi_1 - \frac{1}{2} \phi_2 \right) \right], \qquad
x^{-}  =   \frac{L^2}{2} \left[ t -\frac{3}{4} \left( \psi + 
\frac{1}{2} \be -\frac{1}{2} \phi_1 - \frac{1}{2} \phi_2 \right) \right],
\nonu
\eea
and make a scaling limit around $\rho=0=\theta_1 = \theta_2 =\al$
in the above geometry (\ref{metric}).
By taking the limit $L \rightarrow \infty$ while 
rescaling the coordinates 
\bea
\rho=\frac{r}{L}, \qquad \theta_1 =\frac{ 2 \zeta_1}{\sqrt{3} L}, \qquad
  \theta_2 =\frac{2 \zeta_2}{\sqrt{3} L}, \qquad \al =
\frac{\sqrt{2} \zeta_3}{\sqrt{3} L} \nonu
\eea
with the fact that the quantities in the last two lines of (\ref{metric1}) 
do not contribute,
the Penrose limit of the  $AdS_4 \times V_{5,2}$ becomes 
\bea
ds_{11}^2 & = & -4 dx^{+} dx^{-} +
 \sum_{i=1}^3 \left( dr^i dr^i -r^i r^i dx^{+}
dx^{+} \right) \nonu \\
&& + \frac{1}{4}
\left( d \zeta_3^2 +\zeta_3^2 d\phi_3^2
- 2 \zeta_3^2 d\phi_3 dx^{+} \right) + \frac{1}{4} \sum_{i=1}^2 \left(
  d \zeta_i^2 +\zeta_i^2 d\phi_i^2
+ 2 \sqrt{3} \zeta_i^2 d\phi_i dx^{+} \right)
\nonu \\
&= &  -4 dx^{+} dx^{-} +\sum_{i=1}^3 \left( dr^i dr^i -r^i r^i dx^{+}
dx^{+} \right) \nonu \\
& & + \frac{1}{4}
 \left( d w d \bar{w} 
+  i  \left( \bar{w} d w - w d \bar{w} \right) dx^{+} \right)
+ \frac{1}{4}
\sum_{i=1}^2 \left( d z_i d \bar{z}_i 
-  i  \sqrt{3} \left( \bar{z}_i d z_i - 
z_i d \bar{z}_i \right) dx^{+} \right)
\label{ppwave}
\eea
where we define $\phi_3 = \frac{1}{2} \left(\be -\phi_1 -\phi_2 \right)$
and in the last line we introduce the complex coordinate $w = \zeta_3 
e^{i \phi_3}$ for $\R^2$ 
and a pair of complex coordinates $z_i=\zeta_i e^{i \phi_i}$  for $\R^4$. 
Since the metric has a covariantly constant null Killing
vector $\pa / \pa_{x^{-}}$, it is also pp-wave metric.
The pp-wave has a decomposition of the 
$\R^9$ transverse space into $\R^3 \times \R^2 \times \R^4 $ where
$\R^3$ is parametrized by $r^i$, $\R^2$ by $w$ and $\R^4$ by $z_i$.
The symmetry of this background is the $SO(3)$ rotation in $\R^3$. 
%and
%a $U(1)_1 \times U(1)_2 \times U(1)_3$ symmetry corresponding to 
%the rotations of
%$\R^2 \times \R^2 \times \R^2$. 
In the gauge theory side,  
the $SO(3)$ symmetry corresponds to the subgroup of
the $SO(2,3)$ conformal group.
% and $U(1)_1 \times U(1)_2 \times U(1)_3$  
%rotation 
%charges $J_1, J_2$ and $J_3$ correspond to the linear combination of
%$Q_1$ and $R$, $Q_2$ and $R$, and $Q_3$ and $R$
%respectively where $R$ is the $U(1)_R$ charge of the gauge theory and $Q_1,
%Q_2$ and $Q_3$ are the Cartan generators of the $SU(2)_1 \times SU(2)_2 \times
%SU(2)_3$ global symmetry of the dual superconformal field theory.  
Note that the pp-wave geometry (\ref{ppwave}) in the scaling limit reduces to
the maximally supersymmetric pp-wave solution of
$AdS_4 \times \S^7$ \cite{kow,op} 
%through
%$w = e^{i x^{+}} \widetilde{w}$ and $z_i = e^{-i x^{+}} \widetilde{z_i}$  
\bea
ds_{11}^2 = -4 d x^{+} d x^{-} +\sum_{i=1}^9  dr^i dr^i -
\left( \sum_{i=1}^3 r^i r^i  + \frac{1}{4}  \sum_{i=4}^9 
r^i r^i \right) dx^{+} dx^{+}. 
\nonu
\eea 
The supersymmetry enhancement in the Penrose limit
implies that a hidden ${\cal N}=8$ supersymmetry is present in the 
corresponding subsector of the dual ${\cal N}=2$ superconformal field 
theory.  In the next section, we provide precise description
of how to understand the excited states in the supergravity theory that 
corresponds in the dual superconformal field theory to operators with a
given conformal dimension. 

%%%%%%%%%%%%%%%%%%%%%%%%%%%%%%%%%%%%%%%%%%%%%%%%%%%%%%%%%%%%%%%%%%%%%%%%%%%%%
\section{Gauge Theory Spectrum}
%%%%%%%%%%%%%%%%%%%%%%%%%%%%%%%%%%%%%%%%%%%%%%%%%%%%%%%%%%%%%%%%%%%%%%%%%%%%%

The 11-dimensional supergravity theory in $ AdS_4 \times V_{5,2}$
is dual to the ${\cal N}=2$ gauge theory 
with gauge group
$USp(2N) \times O(2N-1) $ 
with chiral fields $S^i, i=1,2,3,4$ transforming in the 
$\left(2N, 2N-1 \right) $  color representation 
\footnote{We take the convention of $USp$ group such that
$USp(2)=SU(2)$. } and transforming
in the spinor representation ${\bf 4}$ of the flavor group $SO(5)$  
\cite{cddf}.
The Stiefel manifold $V_{5,2}$ is a coset manifold $SO(5)/SO(3)$
where the embedding of $SO(3)$ in $SO(5)$ is the canonical one: 
the fundamental of ${\bf 5}$ of $SO(5)$ 
breaks into ${\bf 5} \rightarrow {\bf 3} + {\bf 1} +{\bf 1}$ under $SO(3)$.
At the fixed point, the chiral superfield $S^i$ has conformal 
weight $1/3$ with $U(1)_R$ charge $1/3$.
%transform as $({\bf 3},{\bf 1})$ 
%and $({\bf 1},{\bf 2})$ under the $SU(3) \times SU(2) 
%$ global symmetry.
We identify states in the supergravity containing both short
and long multiplets with operators in the gauge theory.
In each multiplet, we specify a $SO(5)$ 
representation \footnote{A representation of $SO(5)$ 
can be identified by a Young diagram
and when we denote the Dynkin label $(a_2, a_1)$, the dimensionality of
an irreducible representation is given by $N(a_1,a_2)=
\frac{1}{6} (a_1+a_2+2)(a_1+1)(a_2+1)(2a_1+a_2+3)$.
Although the representation with its dimensionality
is possible, we use the same notation as \cite{cddf} 
for comparison. That is, $SO(5)$ quantum number, $M$ and $N$
characterized by Young tableau $[{\bf M+N}, {\bf M}]$ such that there are 
$2M+N$ boxes.},
conformal weight and $R$-charge.

$\bullet$ {\bf Massless(or ultrashort) multiplets} \cite{cddf}

$1)$ Massless graviton multiplet: 
$
\left[{\bf 0}, {\bf 0} \right], \qquad
 \Delta=2, \qquad R=0
$ 

There exists a stress-energy tensor superfield $T_{\al \be}(x, \theta)$ 
satisfying
the equation for conserved current $D_{\al}^{\pm} T^{\al \be}(x, \theta) 
=0$. 
This $T_{\al \be}(x, \theta)$ expressed as quadratic in chiral field
$S^i$ \cite{cddf} 
is a singlet with respect to 
the flavor group $SO(5)$ and its conformal dimension is 2 with vanishing
$R$ charge.
So this corresponds to the massless graviton multiplet  that 
propagates in the $AdS_4$ bulk. 

$2)$ Massless vector multiplet:
$
\left[{\bf 1}, {\bf 1} \right], \qquad
 \Delta=1, \qquad R=0
$
 
There exists a conserved vector current, a scalar superfield 
$J_{SO(5)}(x, \theta)$,
to the generator of the flavor
symmetry group  $SO(5)$ satisfying the conservation
equations $D^{\pm \al} D^{\pm}_{ \al} J_{SO(5)}(x, \theta) = 
 0 $. 
This $J_{SO(5)}(x, \theta)$ transforms  in the $\left[{\bf 1}, {\bf 1} 
\right]$  representation
of 
the flavor group and its conformal dimension is 1 with vanishing
$R$-charge.
This corresponds to the massless vector multiplet 
propagating in the $AdS_4$ bulk. 
Therefore massless multiplets 1) and 2) saturate
the unitary bound and have a conformal weight related to the 
maximal spin: $\Delta=s_{\mbox{max}}$ where $s_{\mbox{max}}$ is 2, 1
for graviton and vector multiplet respectively. 

$\bullet$ {\bf Short multiplets} \cite{cddf}

It is known that 
the dimension of the scalar operator in terms of energy labels,
in the dual SCFT corresponding
$AdS_4 \times V_{5,2}$ is
\bea
\Delta = \frac{3}{2} + \frac{1}{2} \sqrt{1 + \frac{m^2}{4}} =
\frac{3}{2} + \frac{1}{2} \sqrt{45 + \frac{E}{4} -6 \sqrt{36 +E}} .
\label{delta}
\eea 
The energy spectrum on $V_{5,2}$ exhibits an interesting feature
which is relevant to superconformal algebra and it is given by
\cite{cddf} 
\bea
E =  \frac{32}{9}
\left( 6M^2 + 9N + 3N^2 +12M +6MN - Q^2  \right) 
\label{energy}
\eea
where the eigenvalue $E$ is classified by
$SO(5)$ quantum numbers $M,N$ (totally we have $2M+N$ boxes)
characterized by Young tableau notation
$\left[ {\bf M+N}, {\bf M} \right]$,
and
$U(1)$ charge $Q$: $M, N=0,1,2, \cdots$ and 
$Q=0, \pm 1, \cdots$. 
The $U(1)$ part of the isometry goup of $V_{5,2}$
acts by shifting $U(1)$ charge $Q$. 
The $R$-charge, $R$ is related to $U(1)$ charge $Q$
by $R=2Q/3$. 
Let us take $R \geq 0$. 
One can find the lowest value of $\Delta$ is equal to
$R$ corresponding to a mode scalar with $M=0, N=3R/2$ 
because $E$ 
becomes $16R(R+3)$ and plugging back to (\ref{delta})
then one obtains $\De=R$. 

Thus we find
a set of operators filling out a $\left[\bf \frac{3R}{2}, {\bf 0}
\right]$ 
multiplet of
$SO(5)$.
The condition 
$\Delta=R$ saturates the bound on $\Delta$ from superconformal algebra.
The fact that the $R$-charge of a chiral operator is equal to
the dimension was observed in \cite{bhk}. 

$1)$ Hypermultiplet:
\bea
\left[\bf \frac{3R}{2}, {\bf 0}
\right], \qquad
 \Delta=R.
\nonu
\eea
The information on the Laplacian eigenvalues allows us 
to get the spectrum of hypermultiplets of the theory corresponding to
the chiral operators of the SCFT.
This spectrum was given in \cite{cddf} and the form of operators
is 
\bea
\mbox{Tr} \Phi_{\mbox{c}} \equiv \mbox{Tr} 
\left(S^{t} \Gamma^a S \right)^{3R/2}
\label{chiral}
\eea
where the flavor indices 
are totally symmetrized,  the chiral superfield
$\Phi_{\mbox{c}}(x, \theta)$ satifies $D_{\al}^{+} 
\Phi_{\mbox{c}}(x, \theta) =0$ and the $\Gamma$'s  are the gamma matrices
in five dimensions. 
The complex coordinates
$z^a$ parametrized by a Calabi-Yau four-fold is given by 
$z^a = \mbox{Tr} \left( S^{t} \Gamma^a S \right)$, $a=1,2,3,4,5$.
The hypermultiplet spectrum in the KK harmonic expansions on $V_{5,2}$
agrees with the chiral superfield predicted by the 
conformal gauge theory.
From this, 
 the dimension of $S^i$ should be 1/3 to match the spectrum.
In fact, the conformal weight of a product of
chiral fields equals the sum of the weights of the single components.
This is due to the the relation of $\Delta=R$ satisfied by
chiral superfields and to the additivity of the $R$-charge.

$2)$ Short graviton multiplet:
$
\left[\bf \frac{3R}{2}, 
{\bf 0} \right], \qquad
 \Delta=R+2
$

The gauge theory interpretation of this multiplet is obtained by
adding a dimension 2 singlet operator with respect to 
flavor group into the above chiral superfield 
$ \Phi_{\mbox{c}}(x, \theta)$.
We consider
$
\mbox{Tr} \Phi_{\al \be} \equiv \mbox{Tr} \left( T_{\al \be}  \Phi_{\mbox{c}} 
\right),$
where $T_{\al \be}(x, \theta)$ 
is a stress energy tensor 
and $ \Phi_{\mbox{c}}(x, \theta)$ is a chiral superfield (\ref{chiral}). 
All color indices are symmetrized before taking the contraction.
It is easy to see this composite operator satisfies $D^{+}_{\al} 
\Phi^{\al \be}(x, \theta) =0 $. When $R$-charge vanishes, it leads to
a massless graviton multiplet we have discussed.

$3)$ Short vector multiplet II:
\bea
\left[\bf \frac{3R}{2}+1, 
{\bf 1} \right], \qquad
 \Delta=R+1.
\nonu
\eea
One can construct the following gauge theory object by recognizing 
that the above $SO(5)$ representation can be decomposed into
$\left[{\bf \frac{3R}{2}}, {\bf 0} \right]$ and 
$\left[{\bf 1}, {\bf 1} \right]$, corresponding to
the short vector multiplet II,
$
\mbox{Tr} \Phi  \equiv \mbox{Tr} \left( J_{SO(5)}  \Phi_{\mbox{c}} 
\right),$
where $J_{SO(5)}(x, \theta)$ 
is a conserved vector current transforming in the $\left[{\bf 1},{\bf 1} 
\right]$ representation
of $SO(5)$ flavor group and 
$ \Phi_{\mbox{c}}(x, \theta)$ is a chiral superfield (\ref{chiral}). 
In this case, we have
$D^{+ \al} D^{+}_{\al} \Phi(x, \theta) =0$. 
This multiplet reduces to a massless vector multiplet when $R=0$. 

$4)$ Short gravitino multiplet II:
\bea
\left[\bf \frac{3(R-1)}{2}, {\bf 0} \right], \qquad \Delta = R+3/2.
\nonu
\eea
The corresponding gauge theory operator
is identified with $\mbox{Tr} \Phi_{\al} \equiv \mbox{Tr} \left[ X_{\al}
\left(S^{t} \Gamma^a S \right)^{3(R-1)/2} \right]$ where
$X_{\al}(x, \theta)$ is a semi-conserved current 
transforming as a singlet of flavor group and satisfying the condition
$D^{+\al} X_{\al}(x, \theta)=0$ by interpreting the
$\Delta$ as a sum of $R-1$ and $5/2$. By construction, we have
$D^{+ \al} \Phi_{\al} =0$.
The operator of conformal dimension 
$\Delta =5/2$( which can be interpreted as the sum of 1, $s=1/2$ and $R=1$
according to the definition of a semi-conserved current) with $R$-charge 1
is given by contracting the $SO(5)$ indices \cite{cddf}
\bea
X_{\al} = S \Ga_a S \left( \overline{S} \Ga_b \overline{S}
 D^{+}_{\al} \overline{S} \Ga^{ab} S - 2 \overline{S}
\Ga_b D^{+}_{\al} \overline{S} \overline{S} \Ga^{ab} S \right).
\nonu
\eea
Note that the conformal dimension and $R$-charge of fermionic coordinate
$\theta_{\al}^{+}$ are $\Delta=1/2$ and $R=1$ respectively. Therefore
the counting of $R=1$ for the $X_{\al}$ field is due to the $D^{+}_{\al}$
with the additivity of $R$-charge( the fact that there are
equal numbers of $S$ and $\overline{S}$ 
does not contribute to the $R$-charge counting).  

$5)$ Short gravitino multiplet I:
\bea
\left[\bf \frac{3R+1}{2}, {\bf 1} 
\right], \qquad \Delta = R+3/2.
\nonu
\eea
In this case, one can think of the following 
gauge theory object corresponding this multiplet
as $ \mbox{Tr} \Psi_{\al} \equiv
\mbox{Tr} \left[ L_{\al}
\left(S^{t} \Gamma^a S \right)^{3(R-1)/2} \right]$
where $L_{\al}(x, \theta)$ is a semi-conserved current transforming
 in $\left[ {\bf 2}, {\bf 1} \right]$ representation of flavor group with
$D^{+\al} L_{\al}(x, \theta)=0$ by writing the $SO(5)$ representation 
in terms of $\left[{\bf \frac{3(R-1)}{2}},{\bf 0} \right]$ 
and $\left[{\bf 2},{\bf 1} \right]$. 
Similarly, one has 
$D^{+ \al} \Psi_{\al}(x, \theta) =0$. The explicit form for 
the operator of conformal dimension $\Delta =5/2$ with $R$-charge 1
being consistent with the definition of semi-conserved current is given by
\cite{cddf}
\bea
\left( L_{\al} \right)^{a(bc)} = \overline{S} \Ga^a S
 D^{+}_{\al} \overline{S} \Ga^{bc} S - 
D^{+}_{\al} \overline{S} \Ga^a S \overline{S} \Ga^{bc} S. 
\nonu
\eea
Therefore the short $OSp(2|4)$ multiplets 1), 2), 3), 4) and 5) 
saturate the unitary bound and have a 
conformal dimension related to the $R$-charge and maximal spin:
$\Delta = R +s_{\mbox{max}}$ where $s_{\mbox{max}}$ is 2, 3/2 and 1
for graviton, gravitino and vector multiplet respectively.  

$\bullet$ {\bf Long multiplets} \cite{cddf}

Although the dimensions of nonchiral operators are in general irrational,
there exist special integer values of $m, n$ such that
for 
\bea
\left[{\bf M+N}, {\bf M} \right]=\left[ {\bf m+n+\frac{3R}{2}},
{\bf m} \right],
\nonu
\eea 
one can see the Diophantine \cite{cddf} like 
condition(See also \cite{ahnplb,ahn02,ahn02-1}), 
\bea
m^2-n^2  -2 mn -3n-m=0
\label{dequation}
\eea
make $\sqrt{36+E}$ be equal to
$4R+2(2m +2n +3)$. 
Furthermore in order to make 
the dimension be rational(their conformal dimensions are protected), 
$45 + E/4 -6 \sqrt{36 +E}$  in (\ref{delta}) should be square of 
something. It turns out this is the case without any further restrictions on
$m$ and $n$. Therefore we  have $\De=R+ m +n$. 
This is true if we are describing states with finite $\Delta$ and $R$.
Since we are studying the scaling limit $\Delta, R \rightarrow \infty$,
we have to modify the above analysis.
This constraint (\ref{dequation}) comes from the fact that the 
energy eigenvalue of the Laplacian on $V_{5,2}$ for the supergravity mode
(\ref{energy})
takes the form 
\bea
E = \frac{32}{3} \left[
5 m^2 +  n^2 + \left( 6 R+7 \right)m + 3 \left( 1 + R \right) n
+ 4mn + \frac{3}{2} R \left( R + 3 \right) \right].    
\label{energyr}
\eea
One can show that the conformal weight of the long vector multiplet II
below becomes rational if the condition (\ref{dequation}) is satisfied.  

$1)$ Long vector multiplet II:
\bea
\Delta= -\frac{3}{2} +\frac{1}{4} \sqrt{E+36}.
\nonu
\eea
However, as we take the limit of $R \rightarrow \infty$, 
this constraint (\ref{dequation}) is relaxed. The combination of 
$\Delta-R$ is given by   
\bea
\Delta-R = m +n  + {\cal O}(\frac{1}{R})
\label{lowest}
\eea
where the right hand side is definitely rational and they are integers.
So the constraint  (\ref{dequation}) is not relevant in the
subsector of the Hilbert space we are interested in. 
Candidates for such states in the gauge theory side are given in terms of
semi-conserved superfields \cite{cddf}. 
Although they are not chiral primaries, their
conformal dimensions are protected. The ones we are interested in take the
following form,
\bea
\mbox{Tr} \Phi_{\mbox{s.c.}} \equiv
\mbox{Tr} \left[  \left( J_{SO(5)} \right)^{m}
  \left( K_{SO(5)} \right)^{n}
\left(S^{t} \Gamma^a S \right)^{3R/2}   \right]
\label{semi}
\eea   
where the scalar superfields $J_{SO(5)}(x,\theta)$ transform 
in the $\left[ {\bf 1}, 
{\bf 1} \right]$ representation of flavor group $SO(5)$ and satisfy
$D^{\pm \al} D^{\pm}_{\al} J_{SO(5)}(x,\theta) =
0 $ with conformal dimension 1
and zero $U(1)_R$ charge. Also 
there exists a scalar superfield $K_{SO(5)}(x, \theta)$ transforming in
$\left[{\bf 1},{\bf 0} \right]$ representation of the flavor group with
same conformal dimension and $R$-charge.
One can construct the following conserved flavor
currents in the $\left[ {\bf 1}, 
{\bf 1} \right]$  representation under $SO(5)$ and
the ones  in the $\left[ {\bf 1}, {\bf 0} \right]$  representation
 as follows:
\bea
 \left( J_{SO(5)} \right)^{ab}  =  
\overline{S} \Gamma^{ab} S, \qquad
 \left( K_{SO(5)} \right)^{a}  =  
\overline{S} \Gamma^{a} S
\nonu 
\eea
where the color indices are contracted in the right hand side.
Note that the conformal dimension of these currents is not
the one of naive sum of $S^i$'s. 
The supergravity theory in $AdS_4 \times
V_{5,2}$ acquires an enhanced ${\cal N}=8$ superconformal symmetry in the
Penrose limit. This implies that the spectrum of the gauge theory 
operators in this subsector should fall into
${\cal N}=8$ multiplets. We expect that both the chiral primary
fields of the form  (\ref{chiral}) and the 
semi-conserved multiplets
of the form (\ref{semi}) combine into  ${\cal N}=8$ multiplets in the 
limit. Note that for finite $R$, the semi-conserved multiplets
should obey the Diophantine constraint (\ref{dequation}) in order for them to
possess rational conformal weights. 
If $m=1$ and $n=0$, we see a shortening of the multiplet related to
the massless vector multiplet or short vector multiplet II depending on
$R=0$ or not.  

In the remaining multiplets we consider the following particular
representations in the global symmetry group $SO(5)$:
$\left[{\bf M+N}, {\bf M} \right]=\left[ {\bf m+n+\frac{3R}{2}},
{\bf m} \right]$.

$2)$ Long graviton multiplet:
\bea
\Delta= \frac{1}{2} +\frac{1}{4} \sqrt{E+36}.
\nonu
\eea
For finite $R$ with rational dimension, 
after inserting the $E$ into the above, 
we will arrive at the relation with same constraint (\ref{dequation})
which is greater than (\ref{lowest}) by 2:
\bea
\Delta-R = 2+ m +n  +{\cal O} \left(\frac{1}{R} \right).
%\label{delta2}
\nonu
\eea
The gauge theory interpretation of this multiplet is
quite simple. If we take  a semi-conserved current 
$\Phi_{\mbox{s.c.}}(x, \theta)$ defined in (\ref{semi})
and multiply it by a stress-energy tensor superfield 
$T_{\al \be}(x, \theta)$ that is a singlet with respect to the flavor group,
namely 
$
\mbox{Tr} \left( T_{\al \be}  \Phi_{\mbox{s.c.}} 
\right)$,
we reproduce the right $OSp(2|4) \times SO(5)$
representations of the long graviton multiplet with right conformal 
dimension.
Also one can expect that other candidate for this multiplet with different 
$SO(5)$ 
representation by multiplying a semi-conserved current with a quadratic
conserved scalar superfield $J_{SO(5)}^2$: 
$\mbox{Tr} \left( J_{SO(5)}^2  
\Phi_{\mbox{s.c.}} \right)$. In this case,
the constraint for finite $\Delta$ and $R$ is
shifted as $m \rightarrow m+2$.
Similarly $\mbox{Tr} \left( K_{SO(5)}^2  
\Phi_{\mbox{s.c.}} \right)$ with the shift of $n \rightarrow n+2$
and
$\mbox{Tr} \left( J_{SO(5)} K_{SO(5)}  
\Phi_{\mbox{s.c.}} \right)$ with the shift of $m \rightarrow m+1, 
n \rightarrow n+1$.
If $m=0=n$, then the conformal dimension reduces to the shortening
condition related to the protected operator corresponding 
to massless and short graviton multiplets depending on $R=0$ or not.

$3)$ Long vector multiplet I:
\bea
\Delta= \frac{5}{2} +\frac{1}{4} \sqrt{E+36}.
\nonu
\eea 
The combination of $\Delta -R$ with Penrose limit 
$R \rightarrow \infty$ in the gauge theory side becomes
\bea
\Delta-R = 4 + m +n  +{\cal O} \left(\frac{1}{R} \right).
\nonu
\eea
By taking quadratic stress-energy tensor and
multiplying it into a semi-conserved current,
one obtains 
$
 \mbox{Tr} \left(
T_{\al \be}^2 \Phi_{\mbox{s.c.}} \right)$. 
Or one can construct 
$
 \mbox{Tr} \left(
T_{\al \be} J_{SO(5)}^2 \Phi_{\mbox{s.c.}} \right)$
corresponding to
this vector multiplet 
and the constraint coming
from the requirement of rationality of conformal dimension is also changed
for finite $\Delta$ and $R$. 
Also one can do the similar things for 
$
 \mbox{Tr} \left(
T_{\al \be} K_{SO(5)}^2 \Phi_{\mbox{s.c.}} \right)$ and
$
 \mbox{Tr} \left(
T_{\al \be} J_{SO(5)} K_{SO(5)} \Phi_{\mbox{s.c.}} \right)$.
One can see there is no shortening condition for this vector multiplet.

$4)$ Long gravitino multiplet I:
\bea
\Delta= -\frac{1}{2} +\frac{1}{4} \sqrt{E+24}.
\nonu
\eea
There exist special integer values of $m, n$ such that
for $\left[{\bf M+N}, {\bf M} \right]=\left[ {\bf m+n+\frac{3R}{2}},
{\bf m} \right]$, 
one can see the Diophantine like 
condition \cite{cddf}, 
\bea
m^2-n^2  +2 m -2mn=0
\label{dequation1}
\eea
make $\sqrt{24+E}$ be equal to
$4R+2(2m +2n )$. 
It turns out this is the case without any further restrictions on
$m$ and $n$. Therefore we  have $\De=-\frac{1}{2}+ R+ m +n$. 
This is true if we are describing states with finite $\Delta$ and $R$.
Since we are studying the scaling limit $\Delta, R \rightarrow \infty$,
we have to modify the above analysis.
However, as we take the limit of $R \rightarrow \infty$, 
this constraint (\ref{dequation1}) is relaxed. The combination of 
$\Delta-R$ is given by   
\bea
\Delta-R = -\frac{1}{2}+ m +n  + {\cal O}(\frac{1}{R})
%\label{lowest1}
\nonu
\eea
where the right hand side is definitely rational.
So the constraint  (\ref{dequation1}) is not relevant in the
subsector of the Hilbert space we are interested in. 
Although they are not chiral primaries, their
conformal dimensions are protected. The ones we are interested in take the
following form,
\bea
\mbox{Tr} \Phi_{\mbox{s.c.}} \equiv
\mbox{Tr} \left[ L_{\al}  \left(  J_{SO(5)} \right)^{m-1}
  \left(  K_{SO(5)} \right)^{n-1}
\left(S^{t} \Gamma^a S \right)^{3(R-1)/2}   \right]
%\label{semi1}
\nonu
\eea   
by realizing that the conformal dimension  can be written as
$\Delta= 5/2 +(m-1) +(n-1)+ (R-1)$.  
If $m=1=n$, we have short gravitino multiplet I we have seen before.

$5)$ Long gravitino multiplet II:
\bea
\Delta= \frac{3}{2} +\frac{1}{4} \sqrt{E+24}.
\nonu
\eea
The combination of $\Delta -R$
is given by 
\bea
\Delta-R = \frac{3}{2}+ m +n  + {\cal O}(\frac{1}{R}).
%\label{lowest2}
\nonu
\eea
One constructs gauge theory operator corresponding to
this multiplet  as follows:
\bea
\mbox{Tr} \Psi_{\mbox{s.c.}} \equiv
\mbox{Tr} \left[  X_{\al} \left(  J_{SO(5)} \right)^{m}
  \left(  K_{SO(5)} \right)^{n}
\left(S^{t} \Gamma^a S \right)^{3(R-1)/2}   \right]
%\label{semi2}
\nonu
\eea
by writing the conformal dimension $\Delta$ as $5/2 + m+n +(R-1)$.
If $m=0=n$, it is easy to see that there exists a shortening condition:
it leads to a short gravitino multiplet II.

%%%%%%%%%%%%%%%%%%%%%%%%%%%%%%%%%%%%%%%%%%%%%%%%%%%%%%%%%%%%%%%%%%%%%%%%%%%%%
\section{Conclusion}
%%%%%%%%%%%%%%%%%%%%%%%%%%%%%%%%%%%%%%%%%%%%%%%%%%%%%%%%%%%%%%%%%%%%%%%%%%%%%

We described an explicit example of an ${\cal N}=2$
superconformal field theory that has a subsector of the Hilbert space
with enhanced ${\cal N}=8$ superconformal symmetry, in the large $N$ limit
from the study of $AdS_4 \times V_{5,2}$.
The pp-wave geometry in the scaling limit produced to 
the maximally ${\cal N}=8$  supersymmetric pp-wave
solution of $AdS_4 \times \S^7$.
The result of this paper shares common characteristic feature 
of previous case of $AdS_4 \times Q^{1,1,1}$ \cite{ahn02}
and $AdS_4 \times M^{1,1,1}$ \cite{ahn02-1}.
This subsector of gauge theory is achieved by Penrose limit
which  constrains strictly the states of the gauge theory to those
whose conformal dimension and $R$ charge diverge in the large $N$ limit
but possesses finite value $\Delta-R$.
We predicted for the spectrum of $\Delta-R$ of the ${\cal N}=2$
superconformal field theory and proposed how the exicited states in the
supergravity correspond to
gauge theory operators. In particular, both the chiral multiplets 
(\ref{chiral}) and
semi-conserved multiplets (\ref{semi}) of ${\cal N}=2$ supersymmetry should
combine into ${\cal N}=8$ chiral multiplets.
It would be interesting to find out a Penrose limit of other types 
\cite{ahnetal} of M-theory
compactification along the line of \cite{warneretal}. 
These examples have different supersymmetries and the 
structures of four-form field strengths are more complicated than what we 
have discussed so far.         
      
\vskip1cm
$\bf Acknowledgements$

This research was supported 
by 
grant 2000-1-11200-001-3 from the Basic Research Program of the Korea
Science $\&$ Engineering Foundation.

\end{document}